%% file: KensukeHomma.tex
\documentclass{PoS}

\title{Search for critical points by measuring spatial correlation lengths via multiplicity density fluctuations}

\ShortTitle{Search for critical points by measuring spatial correlation lengths}

\author{\speaker{Kensuke Homma} and PHENIX collaboration\\
        Hiroshima University, 1-3-1 Kagamiyama, Higashi-hiroshima, Hiroshima, Japan\\
        E-mail: \email{homma@hepl.hiroshima-u.ac.jp}}

\abstract{\input{abstract.tex}}

\FullConference{The 3rd edition of the International Workshop --- The Critical Point and Onset of Deconfinement --- \\
		 July 3-7 2006\\
		 Galileo Galilei Institute, Florence, Italy}

\begin{document}

\input{intro.tex}
\input{theory.tex}
\input{observable.tex}
\input{analysis.tex}
\input{Fig1.tex}
\input{result.tex}
\input{Fig2.tex}
\input{Fig3.tex}
\input{Fig4.tex}
\input{discussion.tex}
\input{conclusion.tex}

\end{document}

%% file: abstract.tex
The nature of phase boundaries in the QCD phase diagram
has not been satisfactorily explored by experiments. 
Based on the Ginzburg-Landau free energy with a spatially inhomogeneous 
term as a function of a scalar order parameter, it is possible to 
determine spatial correlation lengths from measured two point correlation 
functions in general, where a spatially dependent multiplicity fluctuation
from the mean value is taken as the order parameter. 
The divergence of the correlation lengths can be a robust signature 
of a critical system as well as the disappearance
of $<q\bar{q}>$ condensations. Simultaneous observations of such observables
can probe the nature of the phase boundary conclusively.
In this letter, a present result from the PHENIX experiment will be reported. 
We will focus whether a critical behavior of the phase transition exists or
not by searching for increases of pseudo-rapidity correlation lengths as a 
function of the number of participant nucleons with dominantly low $p_t$ 
charged particles where instant fluctuations such as high $p_t$ jets are 
expected to be negligible. The result shows a non monotonic increase in 
the correlation lengths which can be a symptom of the critical behavior.

%% file: intro.tex
\section{Introduction}
Many observations at RHIC indicate the formation of the strongly interacting 
dense medium with partonic natures\cite{WP}.
Nevertheless, information on the phase boundaries have not been quantified.
Although lattice QCD calculations as well as model calculations
predict the existence of a tricritical end-point,
none of them reaches a quantitative agreement each other
on the location of the end-point\cite{Stephanov:2004wx}.
Hence it is crucial to pin down the end-point in the QCD phase
diagram and investigate the transition order by experiments.
Before pinning down the end-point as an ultimate goal,
it is important to establish ways to determine critical systems in general.
As one of such observables, increases of spatial correlation lengths 
as a function of system energy density can be a robust 
signature to determine critical systems whatever the transition order is.

In the following sections the relation between the phase transition and 
the density fluctuations is briefly reviewed and the experimentally accessible 
observables are advocated. Based on the observables, Au+Au collision events taken
by the PHENIX detector\cite{PHENIXNIM} at $\sqrt{s_{NN}}=200$GeV have been analyzed
and the present results are summarized by focusing on whether critical behaviors of
the phase transition exist or not as a function of the number of participants $N_p$
which reflects the system energy density\cite{Milov}.

%% file: theory.tex
\section{Density fluctuation and phase transition}

%
% Free energy
%
In order to relate the density fluctuations with 
the phase transition in the simplest form,
Ginzburg-Landau(GL)\cite{GL} theory with
the Ornstein-Zernike picture\cite{OZ} for a scalar order parameter 
is briefly reviewed. The first attempt to apply the free energy discussion
to nucleus-nucleus collisions can be found in \cite{Scalapino:1974br}.
GL describes the relation between a free energy density $f$ and
an order parameter $\phi$ as a function of system temperature $T$.
By adding a spatially inhomogeneous term $(\nabla\phi)^2$
and an external field $h$, the general form is described as follows;
\begin{eqnarray}\label{eq1} f(T,\phi,h)= f_0(T)+ \frac{1}{2}A(T)(\nabla\phi)^2+
\nonumber \\ \frac{1}{2}a(T)\phi^2 +\frac{1}{4}b\phi^4+ \cdot\cdot\cdot -h\phi
\end{eqnarray}
where terms with odd powers are neglected due to the symmetry
of the order parameter and the sign of $b$ is used to
classify the transition orders; $b<0$ for first order, $b>0$ for
second order and $b=0$ for the critical point.
Since the order parameter should vanish above critical temperature $T_c$,
it is natural for the coefficient $a(T)$ to be expressed as
$a(T) = a_0(T-T_c)$, while $b$ is usually assumed to 
be constant in the vicinity of $T_c$.

%
% Definition of order parameter
%
In the following analysis, 
the order parameter corresponds to the multiplicity density 
fluctuation from the mean density as a function of one dimensional
rapidity point $y$, which is defined as
\begin{eqnarray}\label{eq2}
\phi(y) = \rho(y) - \langle\rho\rangle
\end{eqnarray}
where a pair of brackets is an operator to take the average.
%
% Free energy deviation
%
With the Fourier expansion of the density fluctuation
$\phi(y) = \sum_k{\phi_k e^{iky}}$ where $k$ is wave number,
one can express the deviation of the free energy density $\Delta F/Y$
due to spatial fluctuations from the equilibrium value as
\begin{eqnarray}\label{eq3}
\Delta F/Y = \frac{1}{Y}\int (f-f_0) dy
= \frac{1}{2}\sum_k{|\phi_k|^2(a(T)+A(T)k^2)}
\end{eqnarray}
where $Y$ is the total rapidity range corresponding to one dimensional volume
and up to the second order terms are taken into account as an approximation
in the vicinity of the critical point in Eq.(\ref{eq1}).
%
% Probability of phi(y)
%
Given the free energy deviation, one can obtain the statistical
weight $w$ for fluctuation $\phi(y)$ to occur in a given temperature $T$
\begin{eqnarray}\label{eq4}
w(\phi(y)) = N e^{-\Delta F/ T}.
\end{eqnarray}
%
% <|phi_k|^2>
%
Therefore the statistical average of the square of the density fluctuation 
with the wave number $k$ is described as
\begin{eqnarray}\label{eq5}
\langle |\phi_k|^2 \rangle = \int_{-\infty}^{+\infty} |\phi_k|^2 
              w\left(\sum_k{\phi_k e^{iky}}\right) d\phi_k \nonumber \\
= \frac{NT}{Y}\frac{1}{a(T)+A(T)k^2}.
\end{eqnarray}

%
% G2
%
Experimentally observable two point density
correlation function can be related to the statistical average of
the square of the density fluctuation.
With a density $\rho(y_i)$ for a given sub volume $dy_i$,
the general two point density correlation $G_2$ is expressed as
\begin{eqnarray}\label{eq6}
G_2(y_1,y_2) = \langle (\rho(y_1)- \langle \rho \rangle )(\rho(y_2)- \langle \rho \rangle ) \rangle
\end{eqnarray}
where the case that $1$ coincides with $2$ is excluded to 
simplify the following discussion. 
Multiplying $e^{-iky} \equiv e^{-ik(y_2-y_1)}$ 
to the both sides of Eq.(\ref{eq6}) 
and integrating over sub volume $dy_1$ and $dy_2$ gives
\begin{eqnarray}\label{eq7}
Y\int G_2(y) e^{-iky}dy = 
\langle | \int (\rho(y)- \langle \rho \rangle ) e^{-iky} dy |^2 \rangle
= \langle |\phi_k|^2 \rangle.
\end{eqnarray}
From Eq.(\ref{eq5}) and (\ref{eq7}), 
$G_2$ can be obtained by the inverse Fourier transformation
of $\langle |\phi_k|^2 \rangle$.
Therefore in the one dimensional case $G_2$ is described as
\begin{eqnarray}\label{eq8}
G_2(y) = \frac{NT}{2Y^2 A(T)} \xi(T) e^{-|y|/\xi(T)},
\end{eqnarray}
where  a correlation length $\xi(T)$ is introduced,
which is defined as
\begin{eqnarray}\label{eq9}
\xi(T)^2 = \frac{A(T)}{a_0(T-T_c)}. 
\end{eqnarray}
In general, a singular behavior of $\xi(T)$ as a function of $T$
indicates the critical point of the phase transition.

%
% Chi
%
The wave number dependent susceptibility can also be
defined from Eq.(\ref{eq1}) and (\ref{eq3}) as follows
\begin{eqnarray}\label{eq10}
\chi_k = 
-\left(\frac{\partial^2 f}{\partial h^2}\right)_T =
\left( \frac{\partial h}{\partial \phi_k} \right)^{-1} =
\left(\frac{\partial^2 (\Delta F/Y)}{\partial\phi_k^2}\right)^{-1} \nonumber \\
= \frac{1}{a_0(T-Tc)(1+k^2\xi(T)^2)} \mbox{\hspace{2cm}}.
\end{eqnarray}
In the case of the static limit of $k=0$, 
the susceptibility can be expressed as
\begin{eqnarray}\label{eq11}
\chi_{k=0} = \frac{1}{a_0(T-Tc)} = \frac{2Y^2}{NT} \xi(T) G_2(0).
\end{eqnarray}
From Eq.(\ref{eq9}) and (\ref{eq11}), simultaneous observations of
singularities or discontinuous behaviors in both $\xi(T)$ and 
$\chi_{k=0} T \propto \xi(T) G_2(0)$ at the same temperature indicate that 
the phase transition is consistent with the second order transition
by assuming continuous variations of $T$.

%% file: observable.tex
\section{Experimental observables}

%
% Connection with F2
%
In the following analysis the density fluctuation is discussed via
the charged particle multiplicity distributions as a function 
of the pseudo-rapidity interval of $\delta \eta$ for each collision centrality.
Let us introduce one and two particle inclusive multiplicity densities
$\rho_1$ and $\rho_2$ based on the inclusive differential cross section 
with respect to the total inelastic cross section $\sigma_{inel}$
as follows\cite{Dremin}
\begin{eqnarray}\label{eq12}
\frac{1}{\sigma_{inel}}d\sigma=\rho_1(\eta)d\eta \nonumber \\
\frac{1}{\sigma_{inel}}d^2\sigma=\rho_2(\eta_1, \eta_2)d\eta_1 d\eta_2.
\end{eqnarray}
With these densities, a two particle density correlation function is defined as
\begin{eqnarray}\label{eq13}
C_2(\eta_1,\eta_2)=\rho_2(\eta_1,\eta_2) - \rho_1(\eta_1)\rho_1(\eta_2).
\end{eqnarray}
The mathematical connection between second order normalized
factorial moment $F_2$ and the two particle correlation function is
expressed as\cite{F2def}
\begin{eqnarray}\label{eq14}
F_2(\delta\eta) = \frac{ \langle n(n-1) \rangle }{ \langle n\rangle^2} =
\frac{\int\!\!\int^{\delta\eta}\rho_2(\eta_1,\eta_2) 
d\eta_1 d\eta_2 }{\{\int^{\delta\eta} \rho_1(\eta) d\eta\}^2} \nonumber \\
=
\frac{1}{(\delta\eta)^2} \int\!\!\int^{\delta\eta} 
\frac{C_2(\eta_1,\eta_2)}{\bar{\rho_1}^2} d\eta_1 d\eta_2 + 1
\end{eqnarray}
where 
$n$ is the number of produced particles,
$\delta\eta$ is the rapidity interval 
which defines the measurable range of $|\eta_1-\eta_2|$,
$\bar{\rho_1}$ is the average number density per unit length
within $\delta\eta$ which is defined as
\begin{eqnarray}\label{eq15}
\bar{\rho_1}=\frac{1}{\delta\eta}\int^{\delta \eta} \rho_1(\eta) d\eta.
\end{eqnarray}

%
% Parametrization with constant term
%
The two particle correlation function $C_2$ can be parametrized
based on the one dimensional function form obtained in Eq.(\ref{eq8}).
However, one has to bear in mind that the damping behavior in Eq.(\ref{eq8})
is originated only from the spatial inhomogeneity of the system 
in a fixed temperature.
In many experimental conditions, the initial system temperature
can not be specified as a point. For instance, corresponding temperature
is indirectly discussed by relating it with the collision centrality.
The centrality bin has a finite size and it causes fluctuations 
originating from the finite temperature bin size. In principle this kind
of fluctuations must be independent of the spatial fluctuations. 
In addition, although a self correlation at the zero distance between 
the two sub volumes in Eq.(\ref{eq6}) was excluded, the self correlation can not be 
excluded in the integrated two particle correlation function contained in
Eq.(\ref{eq14}), since there is no explicit distinction between two particles
in the following analysis procedure.
Therefore the general function form for the normalized two particle correlation 
function in the one dimensional analysis can be parameterized as follows
which explicitly contains a constant term $\beta$;
\begin{eqnarray}\label{eq16}
\frac{C_2(\eta_1, \eta_2)}{\bar{\rho_1}^2} 
= \alpha e^{-\delta\eta/ \xi} + \beta
\end{eqnarray}
where $\bar{\rho_1}$ is proportional to the mean multiplicity in each centrality.

%
% NBD
%
Instead of $F_2$ itself, we will use an indirect parameter $k$ 
of the Negative Binomial Distribution(NBD) in the following analysis
which is defined as 
\begin{eqnarray}\label{eq17}
P_{k,\mu}(n) = 
\frac{\Gamma(n+k)}{\Gamma(n-1)\Gamma(k)} 
\left( \frac{\mu/k}{1+\mu/k} \right) \frac{1}{1+\mu/k},
\end{eqnarray}
where $\mu$ corresponds to the mean value and
$k^{-1}$ reflects deviations from the completely random case $i.e.$
the Poisson distribution which corresponds to $k=\infty$.
Intuitively $k^{-1}$ is a measure how strongly particles are correlated.
The mathematical relation between $k$ and $F_2$ is expressed as\cite{WZ}
\begin{eqnarray}\label{eq18}
k^{-1} = F_2-1.
\end{eqnarray}
The reason why we adopt NBD rather than $F_2$ is that
NBD can provide an approximate probability
distribution which enables us to estimate how inefficient or dead areas of
the detector system bias the $k$ parameter and to obtain the
true value of $k$ based on the estimation, while factorial moment
itself does not provide any specific models on the distribution function
which resulted the observed factorial moment.

%
% Connection to NBD k
%
As the result, the relation between the
NBD $k$ parameter and the pseudo-rapidity interval $\delta\eta$
for the parametrization given in Eq.(\ref{eq16}) is expressed as 
\begin{eqnarray}\label{eq19}
k^{-1}(\delta\eta) = F_2 - 1 =
\frac{2\alpha \xi^2(\delta\eta/\xi-1+e^{-\delta\eta/\xi})}{\delta\eta^2}+\beta.
\end{eqnarray}

%
% Chi*T
%
Once $\alpha$, $\xi$ and $\mu$ are obtained from the NBD fits, 
one can measure the product of the static susceptibility and 
the corresponding temperature based on Eq.(\ref{eq11}) and (\ref{eq16});
\begin{eqnarray}\label{eq20}
\chi_{k=0} T \propto \bar{\rho_1}^2 \xi \alpha \equiv \left(\frac{\mu}{\mu_{max}}\right)^2 \xi \alpha
\end{eqnarray}
where $\mu_{max}$ is the mean multiplicity in the most central collision event sample
in the following analysis. Experimentally it is enough to see how the $\chi_{k=0} T$ 
behaves as a function of a quantity which reflects $T$.

%% file: analysis.tex
\section{Data analysis}
In the data analysis 258k events taken by a minimum bias trigger during Run2
were used. The trigger requires coincident hits 
in Beam-Beam Counters(BBC) and Zero-Degree Calorimeters.
The charged particles were reconstructed by a drift chamber and two multi-wire
chambers with pad readouts without a magnetic field in order to statistically 
enhance the low $p_t$ charged particles. It is essential to focus on the
low $p_t$ charged pions for the discussion of the phase transition.
The charged particles were measured by the east arm of the PHENIX detector
which covers $\pm 0.35$ in pseudo-rapidity and $\pi/2$ in azimuthal angle
$\phi$ around the beam line.
Events with collision points within $\pm 5$cm from the center of the detector
along the beam line were analyzed.
The collision centralities were classified by the correlation between
BBC charge sum and ZDC energy sum\cite{Milov}.
We have checked the detector stability rigorously for a run range 
we have analyzed.

In addition, the dead or inefficient areas in the detector have been
identified and the effects on the NBD parameters were carefully 
studied. We have estimated the relation between input $k$ values of NBD
and biased $k$ values due to dead or inefficient areas
for each size of pseudo-rapidity interval in each centrality.
As long as the baseline distribution is known as NBD, one can generate
the number of particles based on the probability distribution
in all fine $\eta-\phi$ bins respectively.
Dead maps can be produced from the track projection points
in $\eta-\phi$ plane in the real data and the maps were used to 
suppress particles which hit the dead areas defined in the maps.
After applying the dead maps, the biased multiplicity distributions
were fit with NBD in given rapidity intervals again, 
which provides correction factors between the true $k$ values and 
the biased ones.
Therefore systematic errors arise from the definition of the dead maps. 
As a general tendency, the dead maps scales only the absolute values of $k$, 
but not change the correlations between $k$ and 
rapidity interval sizes drastically. 

%% file: Fig1.tex
\begin{figure}
\begin{center}
\includegraphics[width=12.0cm]{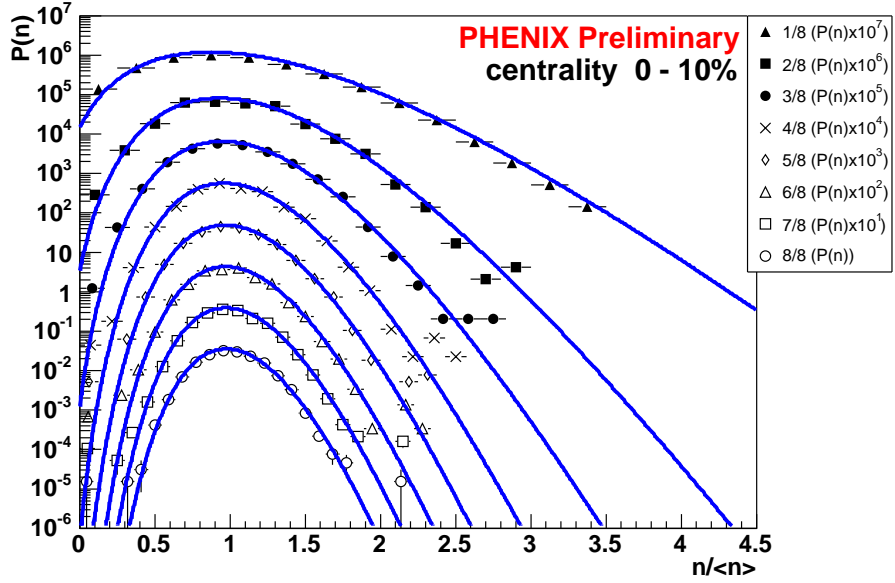}
%\vspace{-0.8cm}
\end{center}
\caption{
Multiplicity distributions in each $\delta\eta$
indicated inside the figure measured in
0-10\% centrality bin in Au+Au collisions at
$\sqrt{s_{NN}}=200$GeV. The horizontal axis
is normalized by the mean multiplicities.
The vertical axis is scaled by the factors
indicated inside the figure.
}
\label{Fig1}
%\vspace{-0.6cm}
\end{figure}

%% file: result.tex
\section{Results} 

Fig.\ref{Fig1} shows the charged particle multiplicity distributions 
in each pseudo-rapidity interval from 1/8 to 8/8 of the full rapidity 
coverage of $|\eta|<0.35$ with 0-10\% events in the collision centrality.
The distributions are shown as a function of the number of
tracks $n$ normalized to the mean multiplicity $\langle n \rangle$.
The vertical error bars show the statistical errors.
The solid curves were determined by performing the NBD fit.
In the following analysis, we have performed the NBD fit
in each pseudo-rapidity interval size from 1/32 to 32/32 of the full rapidity 
coverage of 0.7 to determine a function shape 
in $k$ vs. $\delta\eta$ more precisely.
The mean and RMS of reduced $\chi^2$ values in the NBD fit 
over all centralities and all interval sizes were obtained as
0.69 and 0.72 respectively, which corresponds to typically 95\%
confidence level. Therefore it is good enough to assume NBD
as a baseline multiplicity distribution to obtain the integrated
correlation function through the $k$ parameter based on Eq.(\ref{eq19}).

Fig.\ref{Fig2} shows corrected $k$ parameters as a function of
pseudo-rapidity interval sizes for centrality classes indicated inside the figure.
The upper and lower two panels correspond to 10\% and 5\% centrality bin 
width cases, respectively. 
The vertical error bars show the statistical errors and
boxes show the systematic errors which come from
correction factors on $k$ due to the possible variation of dead or inefficient
areas in the tracking detector. The solid line indicates the
fit result by using Eq.(\ref{eq19}) with errors of quadratic sum of
the statistical and systematic errors. 
The fit was performed from 0.02 to 0.7 in pseudo-rapidity.
The lowest centrality bin was determined as 55-65\%.
The fits are remarkably well resulting reduced $\chi^2$ of 0.44 at the worst
which corresponds to above 99\% confidence level.
This guarantees that the parametrization is actually reasonable.

Fig.\ref{Fig3} a), b) and c) show obtained fit parameters $\alpha$, $\beta$ and $\xi$
as a function of the number of participants $N_p$ where results for
both 10\% and 5\% centrality bin width cases are plotted as red and blue circles respectively.
$N_p$ was obtained from the centrality classes
based on the Glauber model which is explained in \cite{Npart} in detail.
The horizontal errors correspond to ambiguities on the mean values of $N_p$
when the centralities are mapped upon $N_p$.
The vertical error bars are obtained from errors on the fitting parameter by the Minuit program.

Fig.\ref{Fig4} shows the product of static susceptibility and 
corresponding temperature $\chi_{k=0}T$ as a function of the number of participants 
$N_p$ in the case of 10\% centrality bin width. This quantity
is proportional to $\bar{\rho_1}^2 \alpha \xi$ based on Eq.(\ref{eq11}),
where $\bar{\rho_1}^2$ is normalized to 1.0 in the 0-10\% centrality
as defined in Eq.(\ref{eq20}).
The horizontal errors correspond to ambiguities on the mean values of
$N_p$ when the centralities are mapped upon $N_p$.
The errors on $\chi_{k=0} T$ were estimated by taking the
error propagation between the three parameters into account.

%% file: Fig2.tex
\begin{figure}
\begin{center}
\includegraphics[scale=0.9]{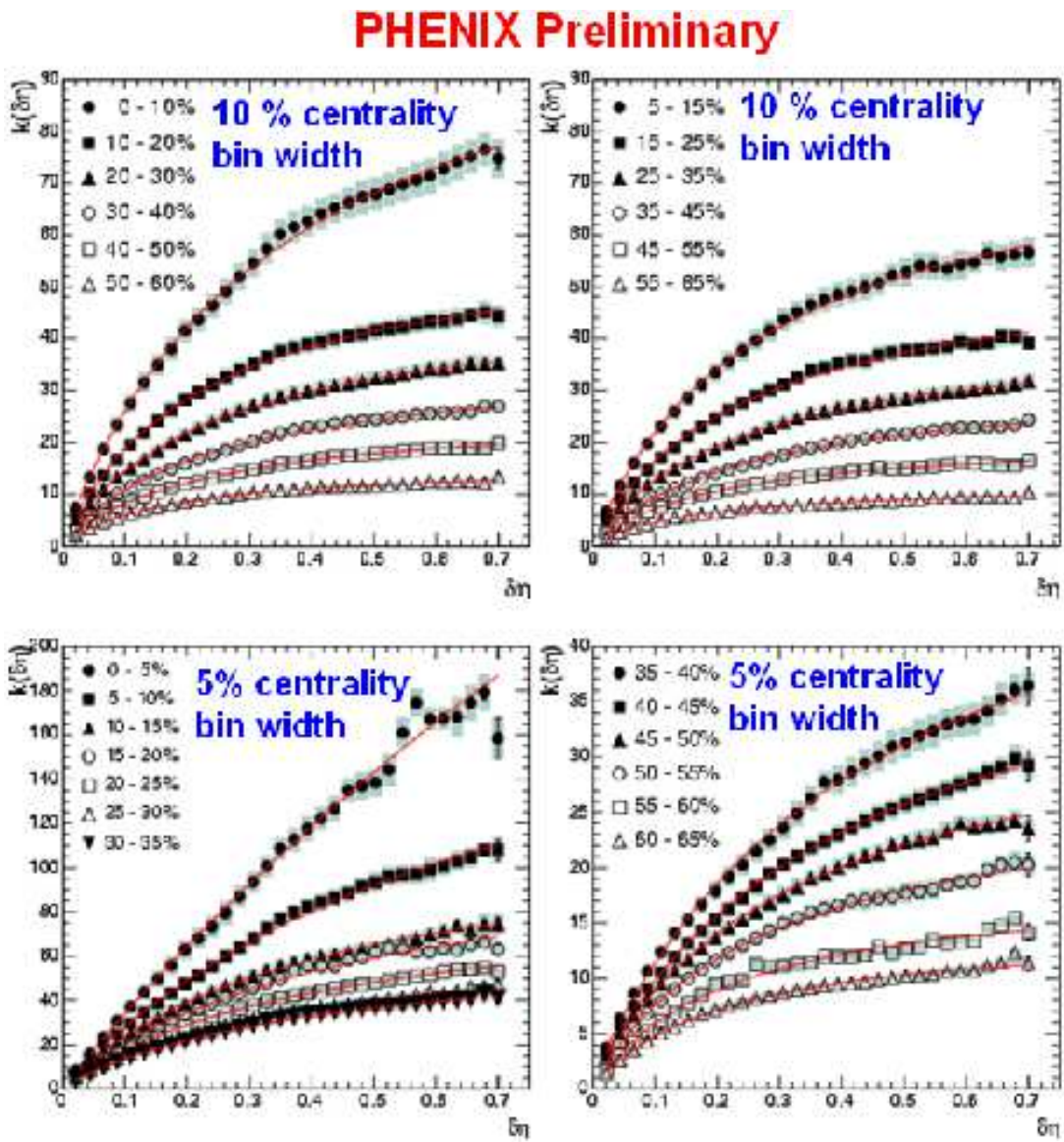}
\end{center}
%\vspace{-0.7cm}
\caption[]{
Corrected $k$ parameters as a function of
pseudo-rapidity interval sizes for centrality classes indicated inside the figure.
The upper and lower two panels correspond to 10\% and 5\% centrality bin
width cases, respectively.
The vertical error bars show the statistical errors and
boxes show the systematic errors which come from
correction factors on $k$ due to the possible variation of dead or inefficient
areas in the tracking detector. The solid line indicates the
fit result by using Eq.(\ref{eq19}) with errors of quadratic sum of
the statistical and systematic errors.
The fit was performed from 0.02 to 0.7 in pseudo-rapidity.
The lowest centrality bin was determined as 55-65\%.
}
\label{Fig2}
%\vspace{-0.3cm}
\end{figure}

%% file: Fig3.tex
\begin{figure}
\begin{center}
\includegraphics[scale=1.0]{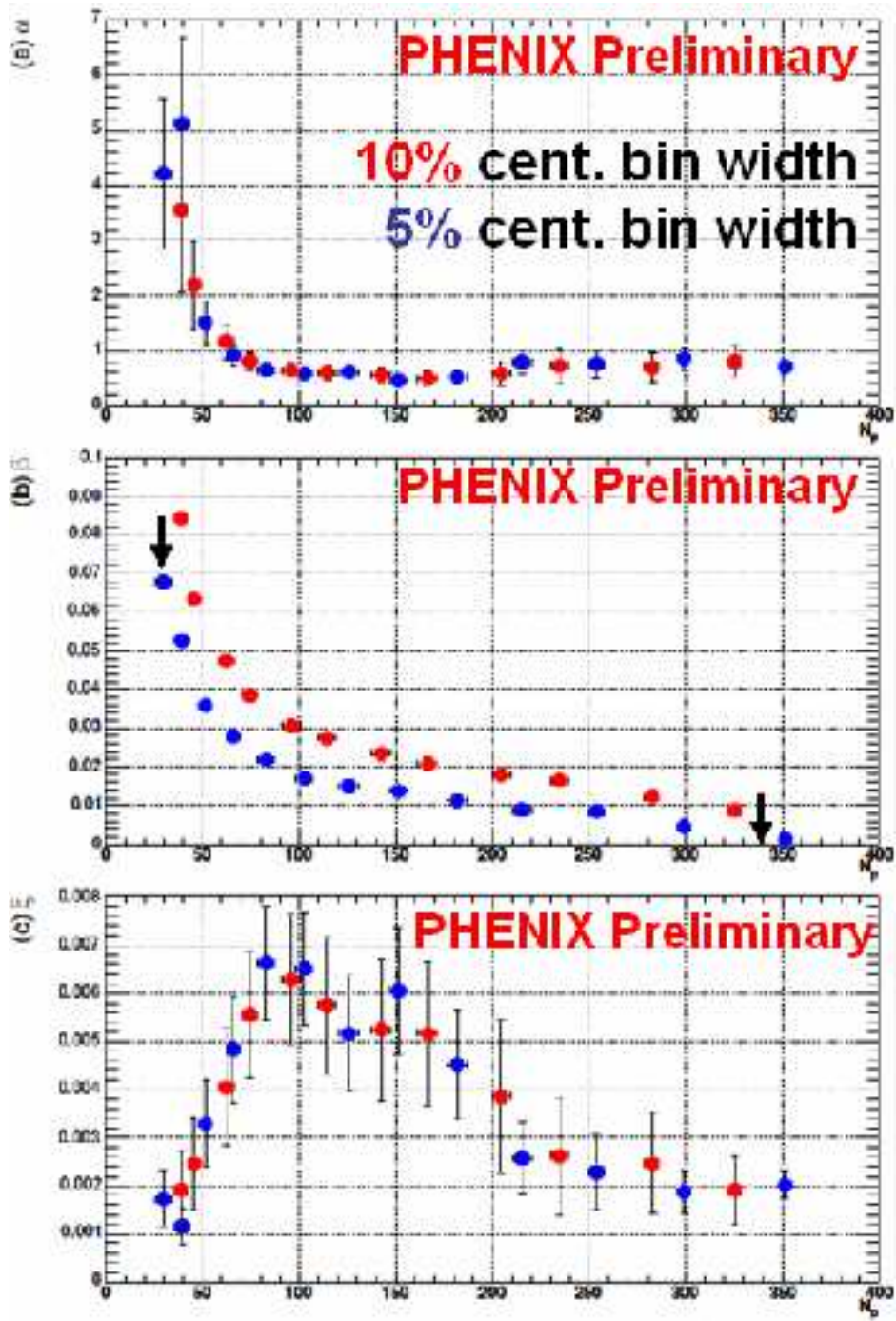}
\end{center}
%\vspace{-0.7cm}
\caption{
Extracted fit parameters $\alpha$, $\beta$ and $\xi$
as a function of the number of participants $N_p$
where results for both 10\% and 5\% centrality bin width cases are 
plotted as red and blue circles respectively.
$N_p$ was obtained from the centrality classes
based on the Glauber model which is explained in \cite{Npart} in detail.
The horizontal errors correspond to ambiguities on the mean values of $N_p$
when the centralities are mapped upon $N_p$.
The vertical error bars are obtained from errors on the fitting parameter
by the Minuit program.
}
\label{Fig3}
%\vspace{-0.5cm}
\end{figure}

%% file: Fig4.tex
\begin{figure}
\begin{center}
\includegraphics[scale=1.0]{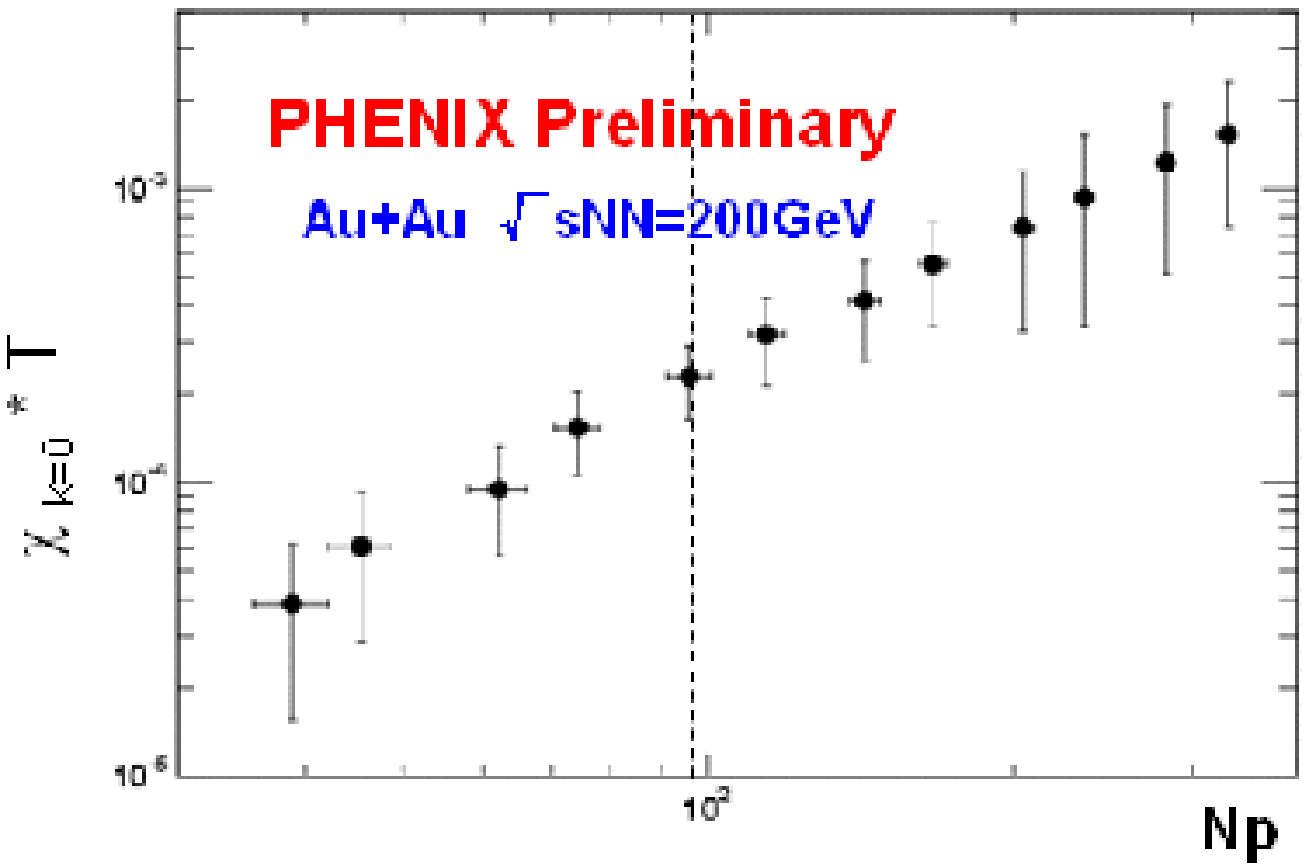}
\end{center}
%\vspace{-0.7cm}
\caption[]{
The product of static susceptibility and
corresponding temperature $\chi_{k=0}T$ as a function of the number of participants
$N_p$ in the case of 10\% centrality bin width. This quantity
is proportional to $\bar{\rho_1}^2 \alpha \xi$ based on Eq.(\ref{eq11}),
where $\bar{\rho_1}^2$ is normalized to 1.0 in the 0-10\% centrality
as defined in Eq.(\ref{eq20}).
The horizontal errors correspond to ambiguities on the mean values of
$N_p$ when the centralities are mapped upon $N_p$.
The errors on $\chi_{k=0} T$ were estimated by taking the
error propagation between the three parameters into account.
}
\label{Fig4}
%\vspace{-0.5cm}
\end{figure}

%% file: discussion.tex
\section{Discussions} 
%
% alpha and shortness of correlation length
%
Since the $\alpha$ parameter corresponds to the correlation strength
at the zero rapidity interval, the correlation is expected to be dominated
by the Bose-Einstein effect in a short range as demonstrated by \cite{MJT2}. 
Since the observed chaoticity parameter $\lambda$ is almost constant as a function of $N_p$
in Au+Au collisions at $\sqrt{S_{NN}}=$200GeV, the assumption of a constant $\alpha$ 
over all centrality samples might be more reasonable rather than 
leaving $\alpha$ as a free parameter within the accessible $\delta\eta$ region.
This is because Eq.(\ref{eq19}) indicates that $k$ can be approximated as 
$1/(2\alpha\xi/\delta\eta + \beta)$ in the limit of $\xi /\delta\eta<<1$ and
there is an inevitable correlation between $\alpha$ and $\xi$ in the limit.
Hence, unless the data points in very small $\delta\eta$ region are available,
one can not completely neglect the correlation between $\alpha$ and $\xi$
in the fit with the integrated two particle correlation function.
The direct measurement of two particle correlation function as demonstrated by
\cite{MJT2} would resolve this issue in future.

%
% role of beta
%
It should be emphasized that the parametrization in Eq.(\ref{eq19}) is robust and
practically works, since the $\beta$ parameter can absorb the effect of centrality 
bin width effects. Fig.\ref{Fig3} b) shows the shift of $N_p$ dependence to lower values 
from 10\% to 5\% centrality bin width case, while other physically crucial parameters
$\alpha$ and $\xi$ are rather stable. In other words,
the $\beta$ parameter contains the effect from the fluctuations on $N_p$.
The ambiguity on $N_p$ measured by PHENIX is not large compared to e.g. 
NA49 where only spectators from the projectile nucleus are measurable 
and it causes an increase of scaled variance of multiplicity distributions in 
peripheral collisions due to dominantly large $N_p$ fluctuations in the target nucleus.
This is due to the partial sampling with respect to the total number of nucleons
in two colliding nuclei. Since both projectile and target nuclei in both sides
can be measured by BBC and ZDC at PHENIX, this kind of large ambiguities on $N_p$ is 
more suppressed even in peripheral collisions. Even if $N_p$ fluctuation is still 
remaining, $\beta$ parameter can absorb this kind of offset parts of fluctuations
and the $N_p$ fluctuation is not harmful for the measurement of correlation lengths 
at all, since correlation lengths are based on the differential of fluctuations.
In addition, $\beta$ would contain effects from the azimuthal correlation such as 
elliptic flows, since the PHENIX does not cover the full azimuth, 
fluctuations caused by reaction plane rotations and elliptic flows should be contained 
in principle. However, since the flow effect looks constant within the experimental 
accuracy in the pseudo-rapidity direction over unit rapidity\cite{RapidityV2},
the measured correlation lengths in the pseudo-rapidity direction
is not expected to be affected by elliptic flows or initial geometrical biases
owing to the $\beta$ parameter.

%
% Np vs. T
%
Concerning the relation between $N_p$ and temperature, it is natural
to assume that $N_p$ can be a monotonically and continuously increasing 
function of the initial temperature at least in the mid-rapidity region, 
since $dE_T/d\eta$ per participant pair is slowly raising as a function 
of $N_p$ as already observed in several RHIC energies\cite{Milov}.
The increase of the correlation length is seen at $N_p\sim 100$ and 
the corresponding energy density based on the Bjorken picture is 
$\epsilon_{Bj}\tau \sim 2.5$GeVfm${}^{-2}$ which has been measured by PHENIX\cite{Milov}.
It is interesting to note that the energy density coincides with the one
where the first drop of $J/\psi$ suppression from the normal nuclear absorption
was observed at SPS\cite{JYSupp}, though the consistency of $E_T$ scales between
two experiments must be carefully checked.

%% file: conclusion.tex
\section{Summary}
The multiplicity distributions measured in Au+Au collisions at 
$\sqrt{s_{NN}}=200$GeV are found to be
well described by the negative binomial distribution.
The two point correlation lengths have been extracted based on 
the function form by relating pseudo-rapidity density fluctuations and
the Ginzburg-Landau theory up to the second order term in the free energy 
with the scalar order parameter. The function form can fit $k$ vs. 
$\delta\eta$ in all centralities remarkably well.
The correlation lengths as a function of the number of participants $N_p$
indicate a non monotonic increase at around $N_p=100$.
This could be a symptom of a critical behavior. Within the present systematic errors,
the product of the static susceptibility $\chi_{k=0}$ and the corresponding
temperature $T$ does not show an obvious turning point
at the same $N_p$ where the correlation length increases.